\documentclass{article}
\usepackage{amsmath}
 \usepackage{amsfonts}
\usepackage{graphics}
\usepackage{graphicx}
\usepackage{geometry}
\begin{document}

\title{Effects of the anomalous
magnetic moments of the quarks on the neutral pion properties
within a SU(2) Nambu-Jona Lasinio model } 

\author{R. M. Aguirre\footnote{e-mail: aguirre@fisica.unlp.edu.ar}
}

\date{Facultad de Ciencias Exactas, Universidad Nacional de
La Plata and  Instituto de Fisica La Plata, Argentina\label{addr1}
}

\maketitle

\begin{abstract}
The properties of the neutral pion in quark matter under the
influence of an external magnetic field are studied. The effects
of the anomalous magnetic moments (AMM) of the quarks at finite
density is considered.  The inclusion of  the AMM into the NJL
model gives rise to additional magnetic effects. In particular the
Dirac sea produce new divergences in the vacuum contributions,
which depend explicitly on the magnetic field. An improper
treatment of these contributions is the source of unphysical
results, as emphasized in recent investigations. The pion
polarization function is evaluated in the random phase
approximation using analytic regularization and a subtraction
scheme to deal with such divergencies. This procedure is combined
with the standard three momentum cutoff, and reduces to it for
vanishing magnetic intensity. The pion mass and coupling constant
are evaluated for a wide range of magnetic intensity and baryonic
density.
\end{abstract}

\section{Introduction}
\label{intro} The study of the strong interaction under the
influence of different  external agents will eventually shed light
on the complexity of its phase diagram, or could reveal new
unknown features. In particular the analysis of the low energy
regime resorts to specific tools such as lattice simulation and
effective models, due to the intricacies of the fundamental
theory. Within the last mentioned case, the Nambu-Jona Lasinio
(NJL) model has shown to be a useful conceptual recourse to tackle
different problems. It has been extensively used to include the
effects of finite temperature, matter density, isospin composition
and
 electromagnetic fields in strong interacting systems
\cite{VOGL,KLEVANSKY,HATSUDA,BUBALLA}. As a special chapter of
this approach one can mention the study of quarks interacting with
external magnetic fields
\cite{MIRANSKY,ANDERSEN,GUSYNIN,EBERT,FERRER,NORONHA,FROLOV,CHATTERJEE,DENKE,FERRER2,AMAN}
, where a variety of issues have been analyzed such as magnetic
catalysis \cite{GUSYNIN}, magnetic oscillations \cite{EBERT},
color superconductivity \cite{FERRER,NORONHA,CHATTERJEE,AMAN},
chiral density waves \cite{FROLOV}, vector \cite{DENKE} and tensor
\cite{FERRER2} additional couplings. The relation to experimental
observables has also been considered, as for instance the dilepton
rate production \cite{ISLAM} and the scattering cross section of
charged mesons \cite{CHANDRA}.
\\
Theoretical treatments based on the local field theory are
compelled to give a proper definition of the vacuum contributions
under the presence of an external magnetic field
\cite{BLAU,GOYAL,ANDERSEN1,COHEN,RUGGERI}.
 Ref. \cite{BLAU} provides general expressions for the effective action of a Dirac
field interacting with a magnetic field for intensities $B$
greater than the mass scale. A description based on the chiral
sigma Lagrangian has been made in \cite{GOYAL}, \cite{ANDERSEN1}
uses the quark-meson model, whereas in \cite{COHEN} the vacuum
contribution to the magnetization is evaluated in a one-loop
approach to QCD for very intense magnetic fields. Using the chiral
quark model and a Ginzburg-Landau expansion Ref. \cite{RUGGERI}
has found that different treatments of the divergences could yield
important modifications of the phase diagram.\\
This discussion is particularly significative for the NJL model
since its main feature is the breakdown of the Lagrangian chiral
symmetry through the vacuum condensates. Therefore the magnetic
effect on the model has been widely debated
\cite{KLEVANSKY,GUSYNIN,EBERT,COPPOLA,MENEZES,FAYAZ2,ANDERSEN2,AVANCINI0,TAVARES,AVANCINI1,AVANCINI2,FAYAZ,CHAUDHURI}.
The evaluation of the effective potential following the analytic
regularization in terms of the Hurwitz zeta function in Ref.
\cite{EBERT} leads to a divergence depending on the squared
magnetic intensity. To dispose of this singularity, the authors
propose a wavefunction renormalization by associating it to the
pure magnetic contribution to the energy density. The same
divergent term has been recently exhibited in the evaluation of
the thermodynamic potential \cite{FAYAZ2}. Furthermore, in a
recent paper \cite{AGUIRR0} this author have shown that the
inclusion of quark anomalous magnetic moments in a SU(2) NJL model
introduce additional divergences depending on $B^2$.\\
The use of different regularization procedures should not lead to
qualitative discrepancies in the physical observables. However, it
has been found  \cite{AVANCINI1,AVANCINI2} that the use of smooth
form factors instead of a steep cutoff could change drastically
the physical predictions. Particularly Ref. \cite{AVANCINI2}
points out that the key point is to clearly distinguish between
the non-magnetic vacuum contribution from the magnetic one.  A
failure in this point should be the cause of the inadequate
behavior found in different calculations, such as tachyonic poles
in the spectrum of light mesons and unphysical oscillations in
thermodynamical quantities.\\
Among the diverse procedures used to deal with the divergences one
can find softening regulators, as used in \cite{FAYAZ} or sharp
step functions \cite{CHAUDHURI}. In both cases there is an
explicit mixing of the integration momentum variable and the
magnetic intensity. More sophisticated is the treatment given in
\cite{SARKAR}, where analytical regularization is used to separate
singularities as poles of the gamma function. Then the integral
representation of the gamma is used modifying its range of
integration to avoid divergences. A method to write the
thermodynamical potential as the sum of terms, one of which
isolate the divergence and does not depend on B, was presented in
\cite{MENEZES}. The authors proposed that this can be achieved by
subtracting and adding the pressure $P_0$ at zero baryonic density
and finite $B$. In the former case $P_0$  still have the
undesirable divergence and in the latter one it has been
regularized in the 3-momentum cutoff scheme.

An interesting aspect of the dynamics of quarks in a magnetic
field is the discrepancy of their gyromagnetic factors from the
ideal value 2. This issue have been considered for a long time
\cite{SINGH,BICUDO,MEKHFI} and has received renewed interest
recently
\cite{FERRER2,FAYAZ,CHAUDHURI,AGUIRR0,ROBERTS,RISCHKE,MAO}. From a
phenomenological point of view one can take as a reference for the
magnetic moments of the light quarks the prediction of the non
relativistic constituent quark model. In order to adjust the
experimental values of the proton and neutron magnetic moments,
the gyromagnetic ratios $\tilde{g}_u=2 \mu_u/\mu_N=3.7$,
$\tilde{g}_d=2 \mu_d/\mu_N=-1.94$ are obtained within this
approach. Given the constituent mass $M$ and the electric charge
$Q_f=q_f\, e$ of a quark of flavor $f$, its AMM can be estimated
as $a_f=\left(2 \tilde{g}_f\, M/ q_f M_p\right)-1$, where $e$ and
$M_p$ are the proton charge and mass respectively.
\\
The AMM of the quarks arises in close relation to the breakdown of
the chiral symmetry. For this reason the NJL model with zero
current quark mass have been used to study the origin of the AMM
\cite{FERRER2,SINGH,BICUDO}.
 To analyze the feasibility of the dynamical generation of the AMM, in
\cite{SINGH} a one loop correction to the electromagnetic vertex
is evaluated within the one flavor NJL model, obtaining
\[ \tilde{g}\approx \frac{2 M_p}{M}\frac{e_q}{e}\left[1-(M/\Lambda)^2 \log(\Lambda/M)\right]. \]

In the approach of \cite{BICUDO} the AMM is extracted from the low
momentum electromagnetic current written in terms of the kernel of
the Ward identities. Assuming a four momentum cutoff, they find
zero AMM in a one flavor NJL model. However,  by using the two
flavor version, the authors obtain $\tilde{g}_u\approx 3.813, \;
\tilde{g}_d\approx -1.929$, which differ from the phenomenological
expectations by less than $1 \%$. Furthermore, in the same work an
schematic confining potential for only one flavor is considered.
By taking a constituent quark mass $M=330$ MeV, typical of the NJL
model, the magnitude of the predicted AMM is as large as 0.15.\\
Another point of view is developed in \cite{FERRER2}, where the
one flavor NJL model is supplemented with a four fermion tensor
interaction, which induces a condensate in the $\gamma^1\,
\gamma^2$ channel. In this case the intrinsic relation between the
 constituent quark mass and its AMM is explicitly exposed, since
the vacuum condensate which breaks the chiral symmetry is also
responsible for the occurrence of nonzero AMM. As a consequence,
the AMM has a non-perturbative dependence on the magnetic intensity.\\
 The necessity of the AMM of the quarks has been
emphasized in \cite{MEKHFI} in the context of the Karl-Sehgal
formula, which relates baryonic properties with the spin
configuration of the quarks composing them. By stating the
dynamical independency of the axial and tensorial quark
contributions to the baryonic intrinsic magnetism, the AMM of the
quarks are proposed as the parameters that distinguish between
them. Resorting to sound arguments, the author propose
$a_u=a_d\approx 0.38$, $a_s\approx 0.2-0.38$ as significative
values for the AMM for the lightest flavors.

Other investigations have focused on the consequences of a linear
coupling between a phenomenological AMM  and an external magnetic
field \cite{FAYAZ,CHAUDHURI,GONZALEZ}. For instance, \cite{FAYAZ}
analyze the phase diagram of the NJL at finite temperature, with
special emphasis on a possible chiral restoration due to the
non-zero AMM. Furthermore, the possibility of a non linear
coupling of the AMM of the quarks is also considered. In this
model the AMM is related to the quantum correction to the
electrodynamic vertex, and a non-perturbative dependence on the
magnetic field is introduced through the effective constituent
quark mass.  The influence of the AMM on the structure of the
lightest scalar mesons is analyzed in \cite{CHAUDHURI}, while
\cite{GONZALEZ} is devoted to study their effects on neutral and
beta stable quark matter within a bag model.

The aim of the present work is to study how some properties of the
neutral pion are modified in the presence of a magnetic field due
the changes in its internal structure.\\
Similar studies, focusing on different light mesons within several
effective models, have been performed in the last years. For
instance, kaons \cite{MISHRA} and pions \cite{MUKHERJEE} were
studied in effective hadronic models, pions in the linear sigma
model \cite{AYALA,DAS}, and also in the NJL model
\cite{TAVARES,AVANCINI1,CHAUDHURI,WANG,LIU}.\\
Effective quarks are considered in the present work, with medium
dependent masses and nonzero AMM. The calculations are performed
by using a quark propagator which includes the anomalous magnetic
moments and the full interaction with the external magnetic field
\cite{AGUIRR0}. As already mentioned, there is recently published
material covering the same subject \cite{FAYAZ,CHAUDHURI}.
However, the main difference with such works is the treatment of
the vacuum contributions. In the cited references the ultraviolet
divergences  are avoided by restricting the range of the integrals
in a way that depends on the magnetic intensity. In the case of
\cite{FAYAZ} the cutoff parameter $\Lambda$ is inspired by a
covariant 4-momentum scheme $E_{n \, s}(p,B)<\Lambda$, where $E_{n
\, s}(p,B)$ represents the energy of the $n$-th Landau level with
spin projection $s$ along the direction of the uniform magnetic
field. Ref. \cite{CHAUDHURI}, instead, uses a 3-momentum cutoff
$p<\sqrt{\Lambda^2+M^2-E_{n \, s}(0,B)^2}$. In contrast, a
complete analytical regularization is given here, that clearly
separates finite magnetic dependency. Since the regularization
procedure for $B=0$ is an intrinsic part of the NJL model, the
results obtained reduce to the standard
vacuum expressions for vanishing magnetic field.\\

This work is organized as follows. In the next section a summary
of the NJL model complemented with a linear coupling for the AMM
is presented and the regularized expressions for the quark
selfenergy and pion polarizations are presented. Some numerical
results for the pion mass and effective coupling are discussed in
Sec. \ref{sec:2}, and the last section is devoted to drawing the
conclusions.

\section{Effects of the AMM on the vacuum properties in the NJL model}
\label{sec:1}

The SU(2) NJL model extended with an AMM term  has the Lagrangian
density

\[
{\cal L}_{NJL}=\bar{\psi}\left(i
\slash\!\!\!\!D-M_0-\frac{1}{2}\kappa \, \sigma_{\mu\,\nu}
F^{\mu\,\nu}\right)\psi+G \left[\left(\bar{\psi}
\psi\right)^2+\left(\bar{\psi} i \gamma^5 \tau_a
\psi\right)^2\right],
\]

where a summation over color and flavor is implicit and $M_0$
stands for the degenerate current mass which explicitly  breaks
the chiral symmetry. The covariant derivative is written as
$D^\mu=\partial^\mu-i Q e A^\mu/3$, where $Q=$diag$(2,-1)$ and a
uniform magnetic field along the $z$ axis is assumed. The
definition $\sigma_{\mu \nu}=i[\gamma_\mu,\gamma_\nu]/2$ is used
and the  AMM are displayed in the matrix $\kappa=$ diag$
(\kappa_u,\kappa_d)$. Each component can be written in terms of
the dimensionless coefficient $a_f$ as
\begin{equation}\kappa_f=\frac{e\,Q_f}{6 M_v}
a_f\label{AMMEq}\end{equation} where $M_v$ stands for the vacuum
value of the constituent quark mass.

This interaction can be used in the mean field approach to
evaluate the quark condensate that gives rise to the constituent
mass, or to search for mesonic modes in a given quark - antiquark
channel. The quark selfenergy as well as the pion polarization
function can be evaluated using the appropriate quark propagator.
The standard approach uses the Schwinger propagator, for a
quasiparticle state dressed by the interaction with a uniform
magnetic field. In the present work an expanded propagator is used
which includes the coupling to the AMM \cite{AGUIRR0}.
\begin{equation}
G_f(x',x)=e^{i \Phi } \int \frac{d^4 p}{(2 \pi)^4} e^{-i
p^\mu\,(x'_\mu-x_\mu)} \, e^{-p_\bot^2/\beta_f}\left[ G_{f 0}(p)+
\sum_{n,s}(-1)^n G_{f n s} (p)\right] \label{PropP}
\end{equation}
where
\begin{eqnarray}
G_{f 0}(p)&=&\left( \not \! u+M_f\mp K_f\right) \left(1\pm i
\gamma^1 \gamma^2\right) \; \Xi_{0 s},\label{LLL}
\end{eqnarray}
\begin{eqnarray}
G_{f n s}(p)&=&\frac{\Delta_n + s M_f}{2 \Delta_n}\Big\{( \not \!
u\mp K_f+s \Delta_n) \left(1\pm i \gamma^1 \gamma^2\right) L_n(2
p_\bot^2/\beta_f)-( \not \! u\pm K_f-s \Delta_n)
\nonumber\\
&&\times \left(1\mp i \gamma^1 \gamma^2\right) \frac{s
\Delta_n-M_f}{s \Delta_n+M_f} L_{n-1}(2 p_\bot^2/\beta_f)\pm
\left( \not \! u\, i \gamma^1 \gamma^2\pm s \Delta_n-K_f\right)
\not \! v \frac{s \Delta_n- M_f}{p_\bot^2}
\nonumber \\
&&\times \left[ L_n(2 p_\bot^2/\beta_f)-L_{n-1}(2
p_\bot^2/\beta_f)\right]\Big\}
\; \Xi_{n s},  \\
\Xi_{n s}&=&\frac{1}{p_0^2-E_{f n s}^2+i\epsilon}+2 \pi
i\,\delta(p_0^2-E_{f n s}^2)\, n_F(p_0)
 \label{DecomP}
\end{eqnarray}
where
\[
 n_F(p_0)=\frac{\theta(p_0)}{1+e^{(p_0-\mu)/T}}-\frac{\theta(-p_0)}{1+e^{-(p_0-\mu)/T}}\nonumber
 \]

The double signs distinguish between positively (upper) and
negatively (lower) charged states. Furthermore,  the index $s=\pm
1$ describes the spin projection on the direction of the uniform
magnetic field. Eq. (\ref{LLL}) propagates the lowest Landau level
with the unique projection $s=1$ for the $u$ flavor and $s=-1$ for
the $d$ case. The sum over the index $n\geq 1$ takes account of
the higher Landau levels, and the following notation is used
$\beta_f=e |Q_f| B/3$, $K_f=\kappa_f\,B$, $\not \! u=p_0
\gamma^0-p_z \gamma^3$, $\not \! v=-p_x \, \gamma^1-p_y\,
\gamma^2$, $p_\bot^2=p_x^2+p_y^2$, $L_m$ stands for the Laguerre
polynomial of order $m$, and
\begin{eqnarray}
E_{f n s}&=&\sqrt{p_z^2+(\Delta_n-s\,K_f)^2}\nonumber \\
\Delta_n&=&\sqrt{M_f^2+2 n \beta_f} \nonumber
\end{eqnarray}
Clearly $n_F(p_0)$ represents the canonical statistical
distribution function for fermions in thermodynamical equilibrium.
Finally, the phase factor $\Phi=\beta_f(x+x')(y'-y)/2$ denotes the
gauge fixing. \\
In the following subsections  some specific details about the
evaluation of the selfenergy and polarization functions are given.

Due to the presence of a vacuum condensate the quark field
acquires an enlarged constituent  mass, a process that in the
usual mean ield approach is described by
\begin{equation} M_i=M_0-
4 G \langle\bar{\psi}_i \psi_i\rangle \label{QMass}
\end{equation}

The quark condensates can be evaluated using the quark propagator
by using the formula \cite{KLEVANSKY}
\begin{eqnarray}
\langle\bar{\psi}_f\,\psi_f\rangle&= &-i\;\lim_{t'\rightarrow
t^+}\,\text{Tr}\{G_f(t,\vec{r},t',\vec{r})\}, \label{Cond}
\end{eqnarray}
The right hand side is ultraviolet divergent and must be redefined
properly. In \cite{AGUIRR0}, this author presented a result based
on a simplifying assumption. As an improvement of such derivation,
 an exact expression is given here.  Starting with
\begin{eqnarray}
\langle\bar{\psi}_f\,\psi_f\rangle=-i\;\lim_{\epsilon\rightarrow
0^+}\,\int\, \frac{d^4p}{(2\pi)^4} \text{Tr}\left\{ G_f(p)\right\}
\label{EDenP}
\end{eqnarray}
a decomposition
\begin{equation}
\langle\bar{\psi}_f\psi_f\rangle=
\langle\bar{\psi}_f\psi_f\rangle^{\text{vac}}+\langle\bar{\psi}_f\psi_f\rangle^{\text{F}}\label{Part0}\end{equation}
is obtained. The first term in this equation corresponds to the
vacuum contribution and it is generated by the first term of Eq.
(\ref{DecomP}).\\
For practical purposes the integration is split into longitudinal
$d^2p_\|=dp_0 dp_z$ and orthogonal $d^2p_\bot=dp_x dp_y$
components. After integrating over the perpendicular momentum one
obtains
\begin{equation}\langle\bar{\psi}_f\psi_f\rangle^{\text{vac}}=-\frac{i }{4
\pi^3}N_c \beta_f M_f \left({\sum_{n,s}}^\prime \int
\frac{d^2p_\|} {p_0^2-E_{f n s}^2+i\epsilon}-K_f
{\sum_{n,s}}^\prime \frac{s}{\Delta_n} \int
\frac{d^2p_\|}{p_0^2-E_{f n s}^2+i\epsilon}\right) \label{Part1}
\end{equation}
where the primed sum indicates that for $n=0$ only one spin
projection must be considered as already explained. Each of the
terms between parenthesis can be evaluated by making a Wick
rotation and introducing an auxiliary integration to bring the
integrand into an exponential form.  Thus, for instance
\begin{eqnarray}
{\sum_{n,s}}^\prime \int&& \frac{d^2p_\|}{p_0^2-E_{f n
s}^2+i\epsilon}=-i \pi \lim_{\epsilon\rightarrow 0}
{\sum_{n,s}}^\prime \Gamma(\epsilon) \left[\frac{\nu}{(\Delta_n-s
K_f)^2}\right]^\epsilon \nonumber \\
&&=-i \pi \lim_{\epsilon\rightarrow
0}\left[\left(\frac{1}{\epsilon}-\gamma\right)\frac{K_f^2-M_f^2}{2
\beta_f}+ \ln\left(\frac{\Gamma(q_f)}{\sqrt{2
\pi}}\right)+\frac{\beta_f+K_f^2-M_f^2}{\beta_f}
\ln\left(\frac{\nu}{2 \beta_f}\right)+\ln\left(\frac{(M_f\pm
K_f)^2}{\nu}\right)+ O(\epsilon) \right]\nonumber\end{eqnarray}

In this formula $\nu$ represents a regularization scale parameter,
which is fixed at $\nu=M_f$, and the definition
$q_f=(M_f^2-K_f^2)/(2 \beta_f)$ is used. The derivation of the
second term is slightly more complex and is left for the Appendix
A. The final result is
\begin{eqnarray}
{\sum_{n,s}}^\prime \frac{s}{\Delta_n} \int \frac{d^2p_\|}
{p_0^2-E_{f n s}^2+i\epsilon}&=& -i \pi \lim_{\epsilon\rightarrow
0} \Bigg\{\frac{1}{\epsilon}\left(\frac{\pm1}{M_f}+\frac{2
K_f}{\beta_f}\right)\mp \frac{1}{M_f}\left[\gamma+ \ln
\left(\frac{(M_f\pm K_f)^2}{\nu}\right)\right]+\frac{2
K_f}{\beta_f} \ln\left(\frac{\nu}{2 \beta_f}\right)- \nonumber
\\ &&\frac{K_f}{\beta_f} \int_0^1 dx \left[\psi(z)+\psi(z^*)+i x \frac{\psi(z)-\psi(z^*)}{\sqrt{1-x^2}}\right]
+ O(\epsilon)\Bigg\}\label{DerivA}\end{eqnarray}
 where
$z=[M_f^2+(1-2 x^2) K_f^2+ i 2 x \sqrt{1-x^2} K_f^2]/2\beta_f$.
Taking account of the properties of the digamma function $\psi$ on
the complex plane, one can verify that the integrand in the last
line is real and well defined, even for
$x=1$.\\
Inserting the two last results into Eq. (\ref{Part1}) one obtains
\begin{equation}\langle\bar{\psi}_f\psi_f\rangle^{\text{vac}}=\frac{N_c }{(2
\pi)^2} M_f \left[\frac{1}{\epsilon}\left(M_f^2+K_f^2\pm
\beta_f\frac{K_f}{M_f}\right)+ {\cal F}(\beta_f,K_f,M_f)+
O(\epsilon)\right] \label{Part2}
\end{equation}
The main term has a contribution $M_f^3/(2 \pi)^2$ coincident with
the standard result of the NJL model at $B=0$. In addition a
contribution depending on $B^2$ can be found. Using the fact that
the main coefficient is a polynomial of degree two in $B$, a
redefinition of the quark condensate is made before taking the
limit $\epsilon \rightarrow 0$
\begin{equation}\langle\bar{\psi}_f\psi_f\rangle^{\text{reg}}=\langle\bar{\psi}_f\psi_f\rangle^{\text{vac}}
-\left(1+\frac{K_f^2}{2}\frac{\partial^2}{\partial K_f^2}+
\frac{K_f}{2} \beta_f \frac{\partial^2}{\partial\beta_f \partial
K_f}\right)\langle\bar{\psi}_f\psi_f\rangle^{\text{vac}}_{B=0}
\label{Part3}
\end{equation}
This expression is free of divergences and becomes zero for $B=0$.
Since one of the purposes of the regularization is to recover the
phenomenology of the NJL model at zero magnetic intensity, a term
$\langle\bar{\psi}_f\psi_f\rangle^{\text{NJL}}$ corresponding to
the standard result using the 3-momentum cutoff $\Lambda$, is
added to (\ref{Part3}). Thus, the final result is
\begin{eqnarray}\langle\bar{\psi}_f\psi_f\rangle^{\text{reg}}=\langle\bar{\psi}_f\psi_f\rangle^{\text{NJL}}&+&\frac{N_c
}{(2 \pi)^2}M_f \Bigg\{M_f^2+2
\beta_f\ln\left(\frac{\Gamma(q_f)}{\sqrt{2
\pi}}\right)+\left(\beta_f-K_f^2-M_f^2\right)
\ln\left(\frac{\nu}{2 \beta_f}\right)\nonumber \\&+&\beta_f
\left(1\pm \frac{K_f}{M_f}\right) \ln\left(\frac{(M_f\pm
K_f)^2}{\nu}\right)\nonumber\\
 &+& K_f^2 \int_0^1 dx \left[\psi(z)+\psi(z^*)+i
x \frac{\psi(z)-\psi(z^*)}{\sqrt{1-x^2}}\right] \Bigg\}\\
\langle\bar{\psi}_f\psi_f\rangle^{NJL}&=&\frac{N_c}{2
\pi^2}M_f\left[M_f^2
\ln\left(\frac{\Lambda+E_\Lambda}{M_f}\right)- \Lambda
E_\Lambda\right]
\end{eqnarray}
The second term in Eq. (\ref{Part0}) corresponds to the
contribution of the Fermi sea of quarks. It is obtained
straightforwardly by using the second term of Eq. (\ref{Part1}).
At zero temperature it is given by
\begin{eqnarray}
\langle\bar{\psi}_f\psi_f\rangle^{\text{F}}&=&\frac{N_c}{2
\pi^2}\beta_f M_f {\sum_{n,s}}^\prime\frac{\Delta_n-s
K_f}{\Delta_n} \ln\left(\frac{\mu_f+p_{f n s}}{\Delta_n-s
K_f}\right),\label{FermiCondensate}
\end{eqnarray}
Where $\mu_f$ stands for the chemical potential and the Fermi
momentum is given by the condition $\mu_f^2=p_{f n
s}^2+\left(\Delta_n -s K_f\right)^2$. It must be noticed that at
zero temperature the sum in Eq. (\ref{FermiCondensate}) extends
till the maximum value $N$ wich satisfies the condition
$\mu_f^2\geq \left(\Delta_n-s K_f\right)^2$. Furthermore, if only
the baryon number is conserved one must take $\mu_u=\mu_d$.

\subsection{The pion polarization function}

The pion is a Goldstone boson in the NJL model, therefore it does
not have pre-existing dynamics and it properties must be defined
ad-hoc. Mesonic excitations in a given quark-antiquark channel can
be found by using the corresponding polarization insertion
$\Pi(q)$. In particular the meson mass $m$ is defined as the
solutions of the equation \cite{VOGL,HATSUDA,KLEVANSKY}
\begin{equation} 1-2 G \Pi(q^2=m^2)=0 \label{PionMass}\end{equation}
Usually the polarization is evaluated in the random phase
approximation which sums up the ring diagram to all orders. Within
this approach the polarization for the neutral pion can be written
as
\[\Pi(q)= i \sum_f \int\, \frac{d^4p}{(2\pi)^4} \text{Tr}\left\{i \gamma_5 G_f(p) i \gamma_5 G_f(p-q)\right\} \]

In this work the pion properties are examined at zero temperature,
for this purpose it is sufficient the quark propagator shown in
Eqs.(\ref{PropP}-\ref{DecomP}), with the fermion distribution
function replaced by the step function $n_F(p_0)=\Theta(\mu-p_0)$.
This propagator has been deduced for positively charged fermions
in \cite{AGUIRRE1} within the  thermal field theory known as
Thermo Field Dynamics \cite{LANDSMAN}. This is a real time
formalism which duplicates its internal degrees of freedom in
order to reproduce the main results and procedures of the field
theory. Therefore the propagator becomes a $2 \times 2$ matrix in
the thermal space, the Eqs.(\ref{PropP}-\ref{DecomP}) describe the
(1,1) component of the matricial arrangement. To extend the
present results to finite temperatures the full matrix structure
must be considered.

 As made for the quark condensate, the polarization can be
decomposed as
\begin{equation}
\Pi(q)=\Pi^{\text{vac}}(q)+\Pi^{\text{F}}(q)\label{Pi0}\end{equation}
where the first term is obtained by using only the first term of
Eq. (\ref{DecomP}) in both propagators. Since the Eq.
(\ref{PionMass}) selects the time-like pion momentum, in the
following ${\boldmath q}=0$ is taken.
This assumption eases the mathematical procedure, although it is not indispensable.\\
A detailed inspection of $\Pi^{\text{vac}}$ shows that after using
the expansion shown in Eq. (\ref{PropP}) for both propagators, the
integration over $p_\bot$ cancels out the crossed terms $G_{f 0}
G_{f n s}$ because of the orthogonality of the Laguerre functions.
The same integration carried out over terms of the form $G_{f n s}
G_{f n' s'}$ turns out proportional to $\delta_{n n'} \delta_{s
s'}$. It deserves to be pointed out that even in the presence of
the AMM the pion vertices do not mix different spin projections.\\
After these simplifications, one can write
\begin{equation}
\Pi_f^{\text{vac}}(q_0)=-\frac{i }{\pi}N_c
\beta_f{\sum_{n,s}}^\prime \left( \int \frac{d^2p_\|} {p_0^2-E_{f
n s}^2+i\epsilon}-\frac{q_0^2}{2}  \int d^2p_\|\left(p_0^2-E_{f n
s}^2+i\epsilon\right)^{-1}\left[\left(p_0-q_0\right)^2-E_{f n
s}^2+i\epsilon\right]^{-1}\right)
\end{equation}
The first term between parenthesis have been examined in the
preceding section. The derivation of the second term is delegated
to the Appendix B, and the final result is shown here
\begin{eqnarray}
{\sum_{n,s}}^\prime \int \frac{d^2p_\|} {\left(p_0^2-E_{f n
s}^2+i\epsilon\right)\left[\left(p_0-q_0\right)^2-E_{f n
s}^2+i\epsilon\right]}&=&i \pi \lim_{\epsilon\rightarrow
0}\Bigg\{\frac{1}{\epsilon}\frac{1}{\beta_f}+2
\ln\left(\frac{\nu}{2\beta_f}\right)+\int_0^1
\frac{dx}{q_0^2x(1-x)-\left(M_f\pm K_f\right)^2}\nonumber\\
&-&\frac{1}{2\beta_f}\left[\psi(q_+)+\psi(q_-)+\frac{K_f}{q_0}\frac{\psi(q_+)-\psi(q_-)}{\sqrt{x(1-x)}}\right]
+O(\epsilon)\Bigg\} \label{DerivB}
\end{eqnarray}
where
\begin{equation}q_\pm=\frac{M_f^2-\left[K_f\pm
x(1-x)q_0\right]^2}{2\beta_f}
\label{Warn}\end{equation}.\\
 Collecting these partial results together one obtains
\begin{equation}\Pi^{\text{vac}}(q_0)=-\frac{N_c }{(2
\pi)^2}
\left[\frac{1}{\epsilon}\left(\frac{q_0^2}{2}-M_f^2+K_f^2\right)+
{\cal G}(\beta_f,K_f,M_f,q_0)+ O(\epsilon)\right] \label{Part4}
\end{equation}
The main coefficient presents the same kind of divergence
$(q_0^2-2 M_f^2)/8\pi^2$ as the usual result at $B=0$,
complemented with a contribution explicitly depending on $B^2$.
Applying the regularization program described at the end of the
preceding section, i.e.
\begin{equation}\Pi^{\text{reg}}(q_0)=\Pi^{\text{NJL}}(q_0)+\Pi^{\text{vac}}(q_0)
-\left(1+\frac{K_f^2}{2}\frac{\partial^2}{\partial
K_f^2}\right)\Pi^{\text{vac}}_{B=0}
\end{equation}
one finally obtains
\begin{eqnarray}
\Pi^{\text{reg}}(q_0)&=&\Pi^{\text{NJL}}(q_0)-\frac{N_c}{4\pi^2}\Bigg\{M_f^2+2
\beta_f\ln\left(\frac{\Gamma(q_f)}{\sqrt{2
\pi}}\right)+\left(\beta_f+K_f^2-M_f^2+\frac{q_0^2}{2}\right)
\ln\left(\frac{\nu}{2 \beta_f}\right)+\beta_f
 \ln\left(\frac{(M_f\pm K_f)^2}{\nu}\right)\nonumber \\
&+&\frac{q_0^2}{2}\left[\beta_f \int_0^1
\frac{dx}{q_0^2x(1-x)-\left(M_f\pm K_f\right)^2}+\int_0^1 dx\,
\ln\left(\frac{M_f^2-q_0^2x(1-x)}{\nu}\right)-\frac{1}{2} \int_0^1
dx \left[\psi(q_+)+\psi(q_-)\right] \right]\nonumber \\
&+&K_f \frac{q_0}{4} \int_0^1 dx
\frac{\psi(q_-)-\psi(q_+)}{\sqrt{x(1-x)}}\Bigg\}\\
\Pi^{\text{NJL}}(q_0)&=&-\frac{N_c}{\pi^2} \int_0^\Lambda dp\, p^2
\frac{\sqrt{p^2+M_f^2}}{p^2+M_f^2-q_0^2/4}
\end{eqnarray}
The last piece is the Fermi sea contribution to the polarization,
which at zero temperature is given by
\begin{equation} \Pi_f^{F}(q_0)=\frac{\beta_f}{2\pi^2} {\sum_{n,s}}^\prime \int_0^{p_{f n s}} dp\,
\frac{\sqrt{p^2+M_f^2}}{p^2+M_f^2-q_0^2/4} \label{FermiP}
\end{equation}
Some imaginary terms giving zero contribution within the regime
considered in the present work have been omitted.\\
The pion mass $m$ found as a solution of Eq. (\ref{PionMass})
gives the ground energy of this bound state. A measure of how this
state interacts with its environment can be obtained by the
effective coupling. The strength $g_{\pi q }$ of the coupling of a
pion with a pair quark-antiquark can be obtained as
\cite{KLEVANSKY}
\begin{equation}
g_{\pi q }^{-2}=\frac{\partial}{\partial q_0^2}\Pi\Big|_{q_0=m}.
\end{equation}

\section{Results and discussion}
\label{sec:2}

In this section the effects of the AMM in the mass of the pion and
its coupling to quarks are studied in terms of the magnetic
intensity and the baryonic number density.\\
In the present calculations the NJL parameters corresponding to
the set 2 of Ref.  \cite{BUBALLA} are used, $M_0=5.6$ MeV,
$\Lambda=587.9$ MeV, $G=2.44/\Lambda^2$. Taking as a guide the
prescriptions $\mu_u=\left(4
\mu_p+\mu_n\right)/5,\;\mu_d=\left(\mu_p+4 \mu_n\right)/5$ of the
constituent quark model together with the experimental values of
the proton and neutron magnetic moments one can find a plausible
range for the AMM. For illustrative purposes two sets are
considered in the present work. For the set AMM1 the numerical
values $\kappa_u=0.074 \mu_N,\;\kappa_d=0.127 \mu_N$, expressed in
units of the nuclear magneton, are chosen. Using $M_v=400$ MeV as
given in the present parametrization  \cite{BUBALLA}, the
coefficients $a_f=0.05, 0.16$ are obtained, see Eq. (\ref{AMMEq}).
Motivated by theoretical estimations \cite{BICUDO,MEKHFI}, the
considerably stronger values $\kappa_u=\kappa_d=0.38 \mu_N$ are
included in the alternative set AMM2. They correspond to
$a_f=0.24, 0.48$, which is compatible with the results of
\cite{MEKHFI}.\\
For given values of $B$ and the baryon density, first the quark
masses must be solved in a self-consistent way. The different
electric charge of the $u$ and $d$ flavors inevitably leads to a
breakdown of the isospin symmetry which is enhanced for higher
$B$. This, in turn, produce a mixing of the pion with other meson
states, as for instance kaons and $\eta-\eta'$. To avoid such
complication some studies use an averaged quark condensate and a
degenerate quark mass. That is, Eq.(\ref{QMass}) is replaced by
\begin{equation} M_u=M_d=M_{\text{symm}}=M_0-4 G
\left(\frac{1}{2} \sum_f
\langle\bar{\psi}_f\psi_f\rangle^{\text{reg}}\right)
\label{SymMass}
\end{equation}
To test the effects of using as input for the solutions of
Eq.(\ref{PionMass}) either the symmetrized (\ref{SymMass}) or the
non-degenerate quark mass (\ref{QMass}) both results are compared
in Fig.1 for the case AMM1.  The results correspond to zero baryon
density and a wide range of intensities. For $B<2.5 \times
10^{19}$ G the discrepancy is negligible. For stronger fields the
symmetrized case gives an almost linear response. The highest
difference is achieved at the end of the scale, representing a 5\%
of deviation. Due to this small difference, in the following the
non degenerate flavor approach is adopted since it also will make
evident the need of new physical input. A full treatment including
meson mixing is pending.\\
The behavior of the pion mass as a function of the intensity $B$
is shown in Fig. 2, for zero baryon density and  different values
for the AMM. All the results coincide in the very low intensity
regime because of the regularization prescription. For
intermediate magnitudes $B< 3\times 10^{19}$ G the calculations
with zero AMM (labelled as AMM0) and using the set AMM1 are
comparable, while the AMM2 case is clearly different. For very
strong fields, the AMM0 curve shows a  decrease until $B\simeq 1.2
\times 10^{20}$ G where a slow increase starts. This behavior is
compatible with the predictions of \cite{TAVARES,AYALA}. The
outcomes of the AMM1 and AMM2 sets are definitely decreasing
functions of $B$. A monotonously decreasing pion mass has also
been obtained in the framework of the linear sigma model under the
strong field treatment of \cite{AYALA}, when dressed couplings are
used. This parameters are obtained at the one-loop level using the
Schwinger propagator \cite{AYALA2} and hence they acquire $B$
dependency. However the decreasing rate is lesser than that in the
AMM1 case. The parametrization AMM2 yields the unexpected
combination of increasing quark masses and collapsing pion mass. \\
These qualitative differences can only be ascribed to the input of
quark masses provided for each set of AMM. In Fig. 3 the
dependence on the intensity $B$ is shown for the deviation of the
symmetrized mass (upper panel)
\[ \Delta M=\frac{M_u+M_d}{2}-M_{\text{symm}}
\]
and the flavor dispersion $M_u-M_d$ (lower panel) of the quark
masses. It is evident that the AMM3 case has the greatest rate of
growth of the flavor splitting, together with the more rapid
deviation from the symmetrized value.  An immediate conclusion is
that the behavior of the in-vacuum pion mass strongly depends on
the rate at which the magnetic intensity affects the flavor
symmetry.\\
In the next figure, the effective coupling of the neutral pion to
the quark-antiquark pair is shown in terms of the intensity $B$.
Results from the AMM0 case are in good agreement with the findings
of \cite{TAVARES}. The set AMM3 distinguishes again from the other
cases keeping slightly below its value for $B=0$.

At this point some remarks about the neutral scalar excitation or
$\sigma$ meson are appropriate. In contrast to the pion vertex,
the scalar one allows the mix of different spin projections, which
in the presence of AMM would result in a non-trivial structure.
Furthermore a regularization program in terms of the zeta function
will conclude probably with digamma functions of arguments as
shown in Eq.(\ref{Warn}). Taking  into account that the sigma
meson mass verifies $M^2-m_\sigma^2/4 <0$ for most of the NJL
parametrizations, one can expect that $q_\pm$ crosses from
positive to negative values as $x$ varies in $[0,1]$. This implies
that the digamma function would cross some of its poles as the
integration is carried out. This remark holds even for zero AMM.

The remaining of this section is devoted to the discussion of the
finite baryon density effects in combination with the magnetism.
It is usual to analyze indirectly these effects by introducing the
associated chemical potential $\mu$. However most of the given
range of variation $\mu < M_f$ reproduce the same state of vacuum.
In this work a different strategy is adopted by recursing to the
isospin conservation \cite{SON,FREEDMAN}. Two particular
configurations will be considered, symmetric matter, where the
Fermi sea is populated with equal number of each flavor $N_u=N_d$,
and electrically neutral matter where $2 N_u=N_d$. They are
schematic descriptions of definite physical situations as  found
in heavy ion collisions experiences or in quark stars,
respectively. The simultaneous conservation of baryon and isospin
charges can be expressed by independent chemical potentials for
each flavor. In Fig. 5 the mass of a pion immersed in dense
symmetric matter is shown as a function of the background magnetic
intensity. Three different values of the baryon number density are
considered $n/n_0=0.5, \, 1, \, 1.5$, where $n_0=0.15$ fm$^{-3}$
represents the equilibrium density of the nuclear matter. For very
low $B$, an increase in density yields an enhancement of the pion
mass as expected for calculations with flavor asymmetry
\cite{RUIVO2,RUIVO3}. For the set AMM0 the density effects are
neutralized by the increase of the magnetic intensity, since after
a slow decrease all the curves corresponding to different
densities coalesce into the same behavior. This is not the case
when the AMM are considered. As can be seen in the panels (b) and
(c) as the density grows the decreasing trend in the high density
regime is emphasized. Thus, for $n/n_0=1.5$ the set AMM1 predicts
a collapse of the pion mass for $B\simeq 7\times 10^{19}$ G. As
already seen, the set AMM2 yields a massless pion even at zero
density for strong enough intensities. This threshold intensities
are $B\simeq 6, 5$ and $3.7 \times 10^{19}$ G for $n/n_0=0, \,0.5$
and $1$, respectively. For $n/n_0=1.5$ the curve for the pion mass
ends abruptly near $B\simeq 2.5\times 10^{19}$ G
because the solutions for the quark masses cease to exist.\\
In Fig. 6 the density effects are analyzed for a configuration of
electrically neutral matter. With the motivation of the high
densities achieved in compact quark stars, which satisfy this
requisite, an extended range of densities $n/n_0 <3$ is
considered. The general trend is similar as that described in Fig.
5, but in this case for  given values of $n,\, B$ the pion mass
is slightly above than in the previous case.\\
It is interesting to note that under specific circumstances the
magnetic field acts as an stabilizer for the pion field. It is
well known that the pion becomes unstable as the condition
$M-m/2<0$ is reached. For $B=0$ an increase in density yields an
increase of $m$, which favors the instability. In contrast, for a
given density $m$ decreases with $B$. The situation is sketched in
Fig. 7, corresponding to the $n_d=2\, n_u$ at fixed density
$n/n_0=4$. The dashed line corresponds to the threshold condition
$m=2\, M$, therefore the region above this curve corresponds to
unstable states. As $B$ grows the system leaves this region and
the pion becomes stable. The same behavior is found for the cases
AMM0
(upper panel) and AMM1 (lower panel). \\
The effective coupling of the pion is revisited in Fig.8 as a
function of $B$ but for fixed density $n/n_0=3$ using the $n_d=2\,
n_u$ configuration. The finite density produces drastic changes as
compared to Fig. 4, there is a general enhancement for all the
range of $B$, indicating a stronger correlation in dense matter.
There is a clearly defined low $B$ regime showing quick variations
which are produced by the occupation of different Landau levels in
Eq. (\ref{FermiP}). As the intensity grows the upper Landau levels
are drained and for instance in the AMM0 case, the $u$ flavor is
completely polarized at $B \simeq 1.7\times 10^{19}$ G. This fact
is manifested by the highest peak in the curve. The $d$ flavor
continues partially polarized until $B \simeq 2.5\times 10^{19}$ G
where it is confined to the lowest Landau level. At this point
$g_{\pi q}$ has its last abrupt drop and enters in a slow regime,
first decreasing and finally monotonously increasing. A similar
description holds for the AMM1 set.

\section{Summary and Conclusions}
\label{sec:3}

In this work some basic properties of the neutral pion immersed in
a background magnetic field are studied, when non-zero anomalous
magnetic moments are assumed for the constituent quarks. For this
purpose the pion polarization function is evaluated in the random
phase approximation using the SU(2) version of the NJL model
supplemented with linear couplings between the external field and
the AMM. Recent publications have warned about the use of an
appropriate regularization of the NJL model in the presence of an
external field to avoid spurious results in physical observables.
In the present calculations it has been shown that the AMM gives
rise to additional divergences depending on the magnetic
intensity, which need an adequate treatment. In consequence an
analytical regularization procedure is proposed to isolate
unambiguously this kind of singularities and to eliminate its
effects on the quark condensate and the pion polarization. As an
extra requisite the regularization used converges to the standard
NJL treatment as $B$
goes to zero.\\
The calculations have been made by using a covariant propagator
for the quarks with constituent mass, which takes account of the
full effect of the magnetic field as well as the effect of the
AMM.\\
The regularized model has been used to study the pion mass $m$ and
effective coupling $g_{\pi q}$, taking the magnetic intensity and
the baryon density as variables. The description of dense matter
have been made assuming isospin conservation and taking as
examples two configurations, flavor symmetric and electrically
neutral quark matter. Two sets of phenomenological AMM are chosen
of different magnitude. the weak (AMM1) and strong (AMM2) sets.  A
wide range of magnetic intensities $B<10^{20}$ G and densities up
to three times the normal nuclear matter value were analyzed.\\
The inclusion of the AMM changes the behavior of the pion mass for
strong fields, giving a decreasing trend for sufficiently high
intensity. Under some circumstances this could lead to the
collapse of the pion mass, particularly for the set AMM2. This
result is emphasized as the baryon density is increased. It has
been found that density and magnetism have opposite effects on the
stability of the pion field in dense matter with isospin
conservation. While a density increase favors the instability, the
magnetic intensity has an stabilizing role by decreasing $m$.\\
 The effective pion-quark coupling shows an
increase of the correlation in the strong field regime for the
vacuum in the AMM1 set. The opposite trend was found for the AMM2
case. The pion-quark interaction is strengthened in dense matter,
and the effective coupling $g_{\pi q}$ gives evidence of the
population of the different Landau levels as $B$ is varied. \\
For a restricted set  $B<2 \times 10^{19}$ G, higher than the
intensities produced in heavy ion collision experiments, both AMM1
and AMM2 sets provide acceptable results. However when the full
range of intensities studied here is considered, the case AMM2
shows some features which make it less eligible. \\
This is a first approach that needs improvement, as for instance
the inclusion of the mixing with other meson states in the
calculation of the polarization function.

\section{Appendix A: Derivation of Eq. (\ref{DerivA})}
\label{AppA}

As a first step, the primed summation in the second term on the
right hand side of Eq. (\ref{Part1}) is completed by summing and
subtracting the same term
\begin{eqnarray}
{\sum_{n,s}}^\prime \frac{s}{\Delta_n} \int \frac{d^2p_\|}
{u_p^2-(\Delta_n-s K_f)^2+i\epsilon}&=&\sum_n \int
\frac{d^2p_\|}{\Delta_n}\left[\frac{1} {u_p^2-(\Delta_n-
K_f)^2+i\epsilon}-\frac{1} {u_p^2-(\Delta_n+
K_f)^2+i\epsilon}\right]\nonumber \\&\pm&\frac{1}{M}\int
\frac{d^2p_\|}{u_p^2-(m\pm K)^2+i\epsilon}
\end{eqnarray}
The right hand side can be rewritten after a Wick rotation as
\begin{eqnarray}
-2 i K_f \sum_n \int
\frac{d^2p_E}{u_E^2+\Delta_n^2+K_f^2}\left[\frac{1}
{u_E^2+(\Delta_n- K_f)^2}+\frac{1} {u_E^2+(\Delta_n+
K_f)^2}\right]\mp\frac{i}{\nu M_f}\int
\frac{d^2p_E\;\nu}{u_E^2+(m\pm K)^2}
\end{eqnarray}
where $d^2p_E=dp_z\,dp_4$ and $u_E^2=p_z^2+p_4^2$. Furthermore an
undetermined
scale constant $\nu$ has been introduced.\\
In the first term one can apply the following relation to each
term between square brackets
\begin{equation}\frac{1}{A B}=\int_0^1
\frac{dx}{\left[x A+(1-x) B\right]^2}\label{FTrick}\end{equation}
to obtain
\begin{eqnarray}
-2 i K_f \sum_{n,s=\pm 1} \int d^2p_E \int_0^1
\frac{dx}{\left[u_E^2+\Delta_n^2+K_f^2+2 x s K_f
\Delta_n\right]^2}\mp\frac{i}{\nu M_f }\int
\frac{d^2p_E\;\nu}{u_E^2+(m\pm K)^2}
\end{eqnarray}
At this point an integration over the auxiliary variable $\tau$ is
introduced to bring the integrands into an exponential form, which
allows an easy integration over $d^2p_E$ yielding
\begin{eqnarray}
-2 \pi i K_f \sum_{n,s=\pm 1}\int_0^1 dx \int_0^\infty d\tau
\exp\left[-\tau \left(\Delta_n^2+K_f^2+2 x s K_f
\Delta_n\right)\right]\mp\frac{i \pi}{\nu M_f }\int_0^\infty
\frac{d\tau}{\tau} \exp\left[-\tau(M_f\pm K_f)^2/\nu\right]
\end{eqnarray}
The integration in the first term is easily done and after some
algebra the last equation takes the form
\begin{eqnarray}
-2 \pi i K_f \sum_n &&\int_0^1 \frac{dx}{\nu}
\Bigg\{\frac{\nu}{\left[\Delta_n^2+\left(1-2 x^2+2 i x
\sqrt{1-x^2}\right)
K_f^2\right]^{1+\epsilon}}+\frac{\nu}{\left[\Delta_n^2+\left(1-2
x^2-2 i x \sqrt{1-x^2}\right) K_f^2\right]^{1+\epsilon}}\nonumber
\\&+&\frac{i  x}{\sqrt{1-x^2}}\left(
\frac{\nu}{\left[\Delta_n^2+\left(1-2 x^2+2 i x
\sqrt{1-x^2}\right)K_f^2\right]^{1+\epsilon}}
-\frac{\nu}{\left[\Delta_n^2+\left(1-2 x^2-2 i x
\sqrt{1-x^2}\right)
K_f^2\right]^{1+\epsilon}}\right)\Bigg\}\nonumber
\\&\mp&\frac{i \pi}{M_f}\int_0^\infty d\tau \tau^{-1+\epsilon} \exp\left[-\tau(M_f\pm
K_f)^2/\nu\right]
\end{eqnarray}
After finishing the algebra an infinitesimal parameter $\epsilon
\rightarrow 0$ is introduced in order to  isolate the singularity
of both integrals. Remembering that $\Delta_n^2=2 \beta_f n
+M_f^2$ and using the definitions of the gamma and the Hurwitz
zeta functions, one finds
\begin{eqnarray}
- \pi i \frac{K_f}{\beta_f}\left(\frac{\nu}{2
\beta_f}\right)^\epsilon \int_0^1 dx \left\{\zeta(1+\epsilon, z)+
\zeta(1+\epsilon, z^\ast)+\frac{i
x}{\sqrt{1-x^2}}\left[\zeta(1+\epsilon, z)- \zeta(1+\epsilon,
z^\ast)\right]\right\}\mp \frac{i \pi}{M_f} \Gamma(\epsilon)
\left[\frac{\nu}{(M_f\pm K_f)^2}\right]^\epsilon \nonumber
\end{eqnarray}
where the definition of $z$ is given immediately after Eq.
(\ref{DerivA}).\\
After making a Laurent expansion around $\epsilon=0$ and arranging
according to same order of $\epsilon$ one finally obtains the
desired Eq. (\ref{DerivA}).\\
Use of the relation
\[\zeta(1+\epsilon,z)=\frac{1}{\epsilon}-\psi(z)+O(\epsilon)\]
has been made.

\section{Appendix B: Derivation of Eq. (\ref{DerivB})}
\label{AppB}

 To start the Feynman propagators are replaced by its
Euclidean version and then the relation (\ref{FTrick}) is used to
rewrite the integrand

\begin{eqnarray}
i \;{\sum_{n,s}}^\prime &&\int \frac{d^2p_E} {\left[u_{p
E}^2+\left(\Delta_n-s K_f\right)^2\right]\left[u_{(p-q)
E}^2+\left(\Delta_n-s K_f\right)^2\right]}\nonumber \\&=&i\;
{\sum_{n,s}}^\prime \int d^2p_E \int_0^1 dx \left\{\left[u_{p
E}-(1-x) u_{q E}\right]^2+(\Delta_n-s K_f)^2+x(1-x) u_{q
E}^2\right\}^{-2}
\end{eqnarray}
After introducing an integration over the auxiliary variable
$\tau$ one gets
\begin{eqnarray}
&i& {\sum_{n,s}}^\prime \int d^2p_E \int_0^1 dx \int_0^\infty
d\tau \,\tau\, \exp -\tau\left\{\left[u_{p E}-(1-x)u_{q
E}\right]^2+(\Delta_n-s K_f)^2+x(1-x) u_{q E}^2\right\}\nonumber
\\&=&i \;{\sum_{n,s}}^\prime \int_0^1 dx \int_0^\infty
d\tau \, \exp \left\{-\tau\left[(\Delta_n-s K_f)^2+x(1-x) u_{q
E}^2\right]\right\}= i\,\pi\, {\sum_{n,s}}^\prime \int_0^1
\frac{dx}{(\Delta_n-s K_f)^2+x(1-x) u_{q E}^2} \nonumber
\end{eqnarray}
As the next step a return to the Minkowski space is made, and then
 to modify the summation the same term is added and subtracted
\begin{eqnarray}
 i\,\pi&&\int_0^1 dx \left\{\frac{-1}{(M_f\pm K_f)^2-x(1-x)
u_q^2}+\sum_n\left[ \frac{1}{(\Delta_n- K_f)^2-x(1-x)
u_q^2}+\frac{1}{(\Delta_n+ K_f)^2-x(1-x)
u_q^2}\right]\right\}\nonumber\\
&=&- i\,\pi\,\int_0^1 \frac{dx}{(M_f\pm K_f)^2-x(1-x)
u_q^2}+i\,\pi\,\sum_n \int_0^1 dx
\Bigg\{\frac{1}{\left[\Delta_n^2-\left(K_f+q_0
w\right)^2\right]^{1+\epsilon}}+\frac{1}{\left[\Delta_n^2-\left(K_f-q_0
w\right)^2\right]^{1+\epsilon}}\nonumber \\
&&+\frac{K_f}{w q_0}\left[\frac{1}{\left[\Delta_n^2-\left(K_f+q_0
w\right)^2\right]^{1+\epsilon}}-\frac{1}{\left[\Delta_n^2-\left(K_f-q_0
w\right)^2\right]^{1+\epsilon}}\right]\Bigg\}
\end{eqnarray}
After completing the algebra inside the summation, an
infinitesimal parameter $\epsilon \rightarrow 0$ is introduced in
order to isolate the singularities. Furthermore the abbreviation
$w=\sqrt{x (1-x)}$ was defined. Using that $\Delta_n^2=2 \beta_f n
+M_f^2$ and recursing to the definition of the Hurwitz zeta
function, one finds
\begin{eqnarray}
i&&\frac{\pi}{2 \beta_f} \left(\frac{\nu}{2
\beta_f}\right)^\epsilon\int_0^1 dx
\left\{\zeta(1+\epsilon,q_+)+\zeta(1+\epsilon,q_-)+\frac{K_f}{w
q_0}\left[\zeta(1+\epsilon,q_+)-\zeta(1+\epsilon,q_-)\right]\right\}\nonumber
\\ &&- i\,\pi\,\int_0^1 \frac{dx}{(M_f\pm K_f)^2-x(1-x)
u_q^2}
\end{eqnarray}
Where $q_\pm$ are defined by Eq. (\ref{Warn}).\\
Finally by making a Laurent expansion around $\epsilon=0$ and
collecting terms of the same order in $\epsilon$ one obtains the
desired Eq. (\ref{DerivB}).

\section*{Acknowledgements}
This work has been partially supported by a grant from the Consejo
Nacional de Investigaciones Cientificas y Tecnicas,  Argentina
under grant PIP 11220150100616CO.

\newpage
\begin{figure}
 \includegraphics[width=0.9\textwidth]{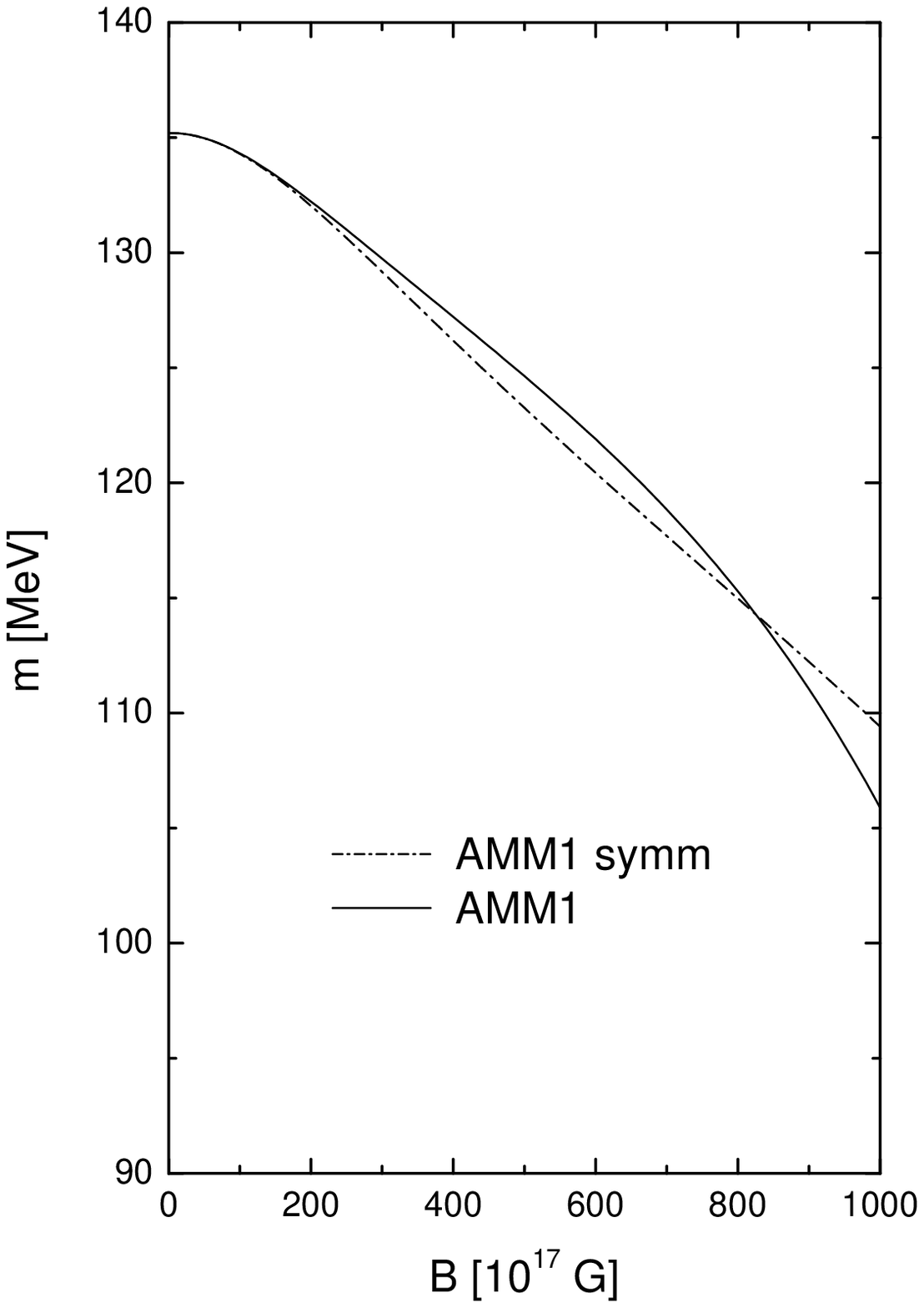}
 \caption{ The in vacuum pion mass using the set AMM1 as a function of the magnetic intensity. Two
 results using as input a degenerate symmetrized quark mass $M_{symm}$ defined by Eq.(\ref{SymMass})
 or the non degenerate $M_u \neq M_d$ case are shown.}
 \end{figure}

\newpage
\begin{figure}
\includegraphics[width=0.9\textwidth]{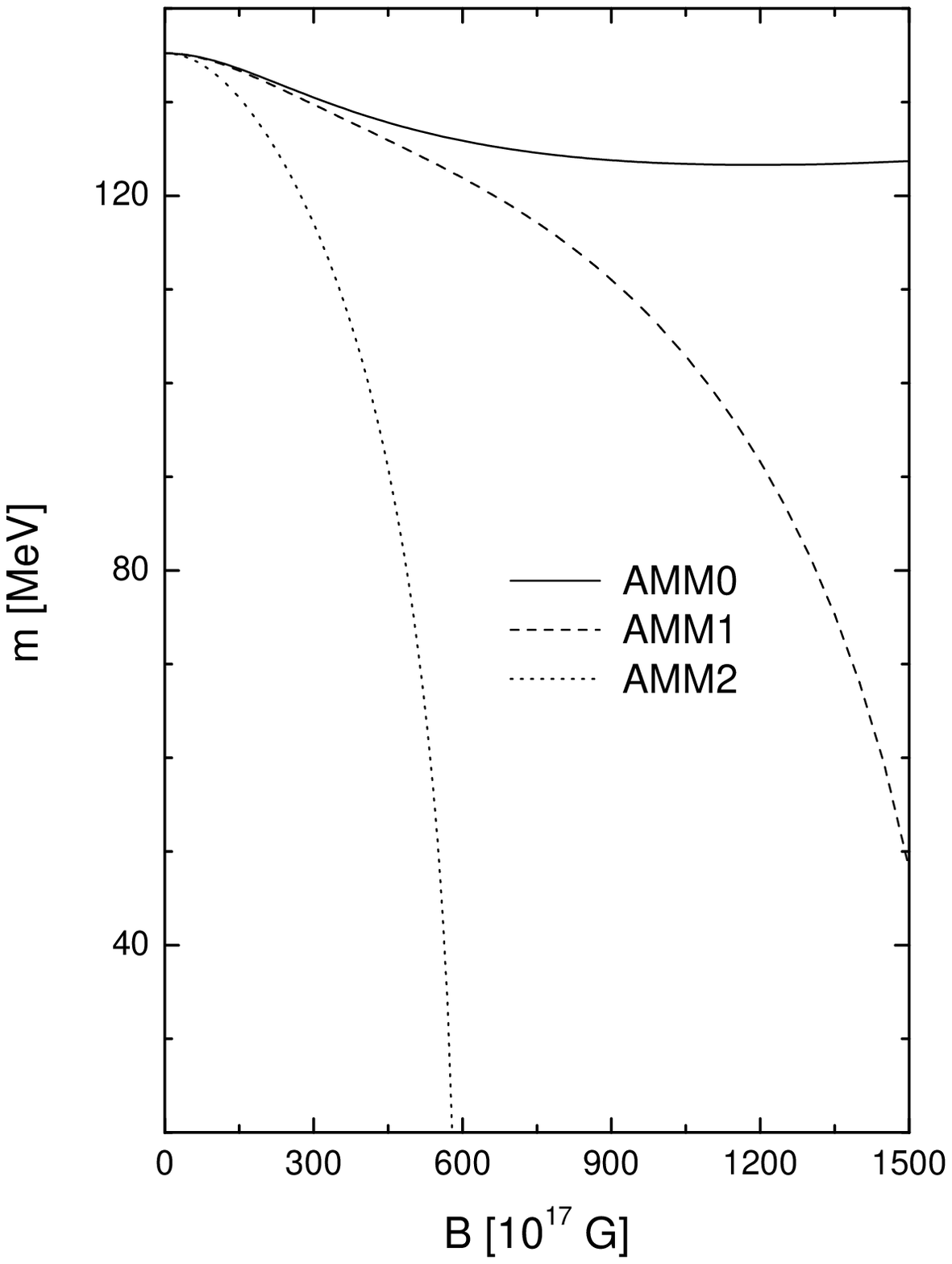}
\caption{ The in vacuum pion mass as a function of the magnetic
intensity. The results without AMM (AMM0) and two different sets
of AMM (AMM1 and AMM2) are compared.}
\end{figure}

 \newpage
 \begin{figure}
 \includegraphics[width=0.9\textwidth]{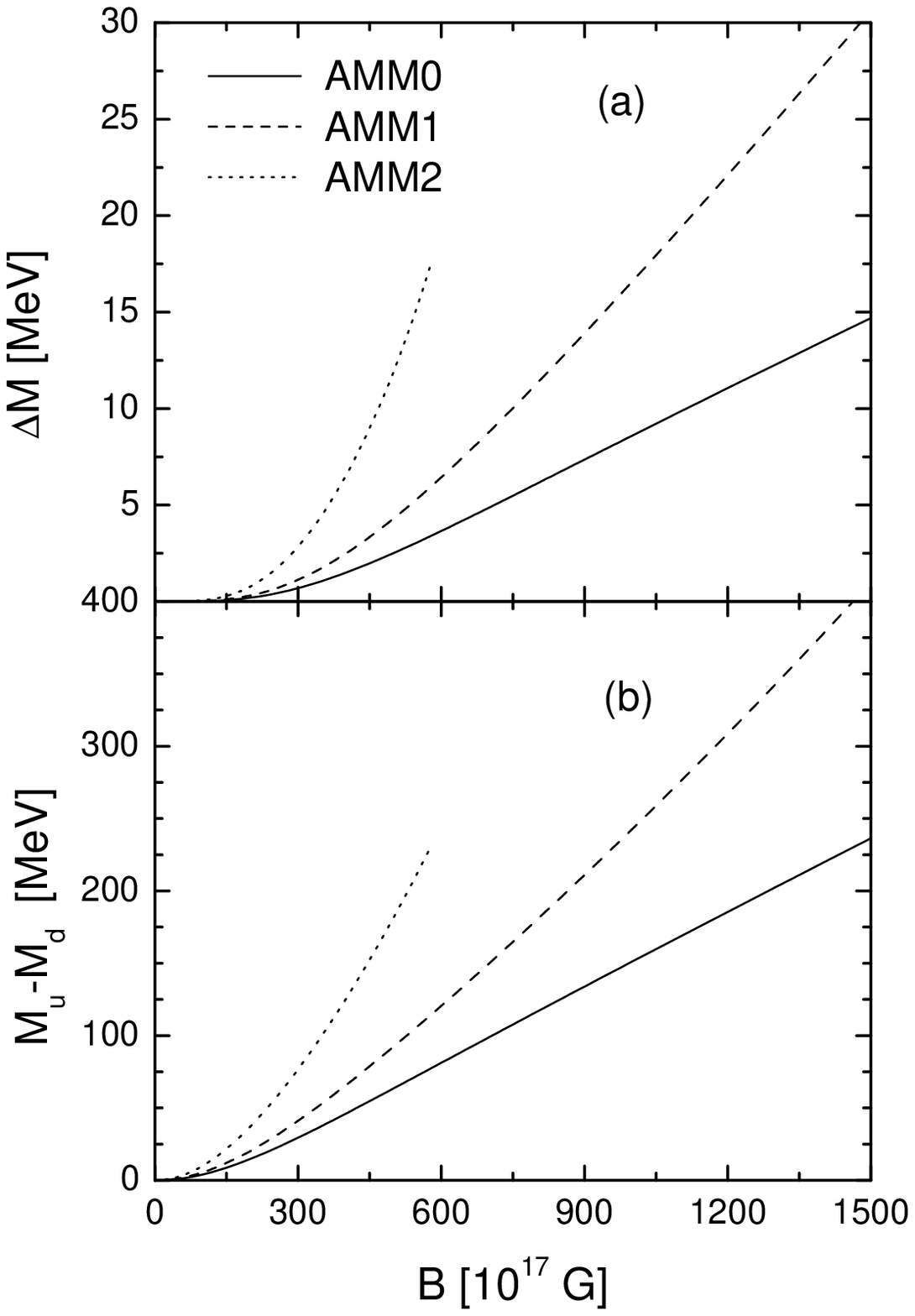}
 \caption{ Two parameters characterizing the inputs used in the pion polarization function
 as functions of the magnetic intensity. In the upper panel (a) the deviation of the average
 $(M_u+M_d)/2$ from the symmetrized $M_{symm}$ values, in the lower panel the flavor splitting
 $M_u-M_d$ are shown.}
 \end{figure}

\newpage
\begin{figure}
\includegraphics[width=0.9\textwidth]{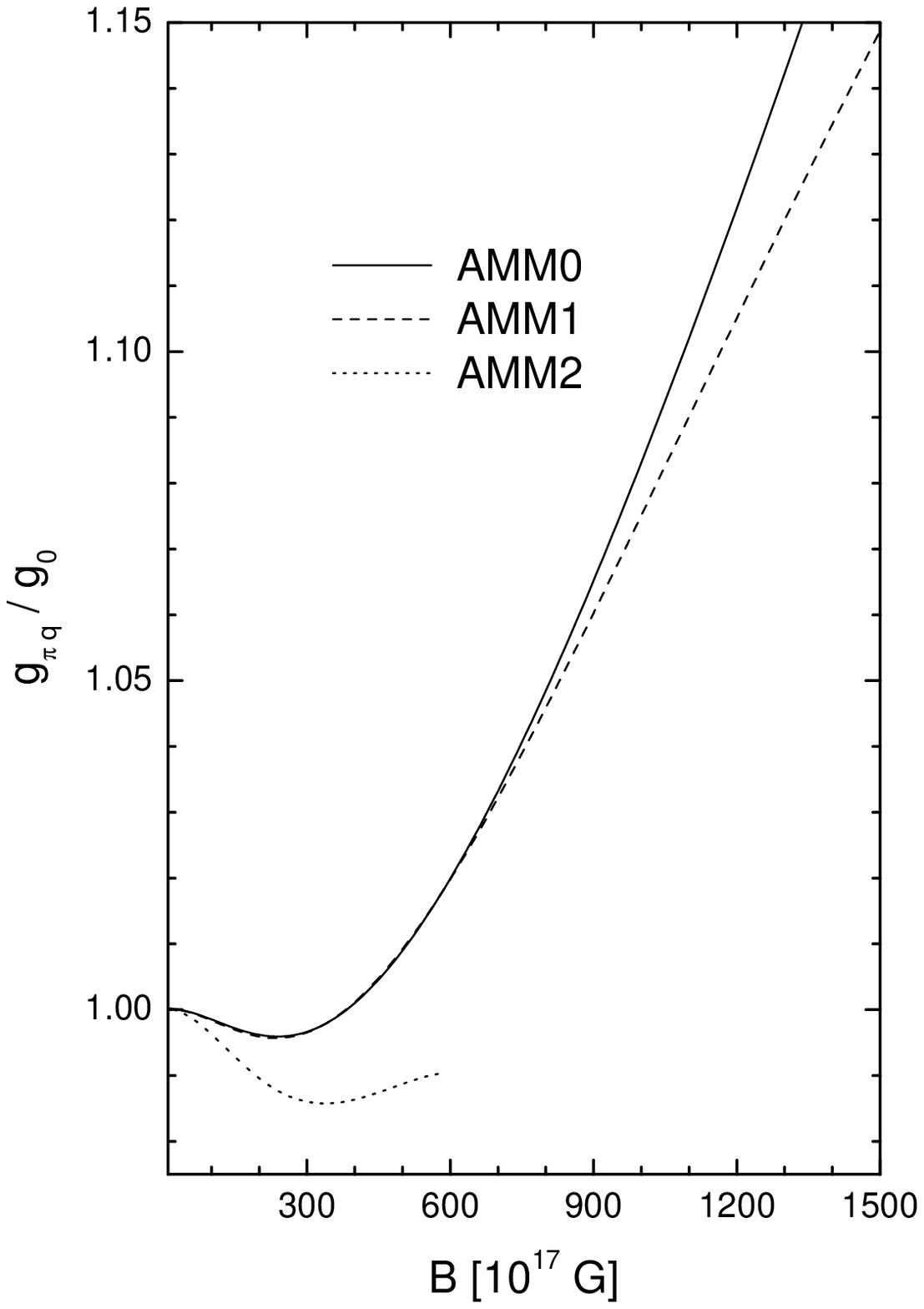}
\caption{ The pion-quark effective coupling as a function of the
magnetic intensity evaluated using three different sets  AMM0,
 AMM1  and AMM2. }
\end{figure}

\newpage
\begin{figure}
\includegraphics[width=0.9\textwidth]{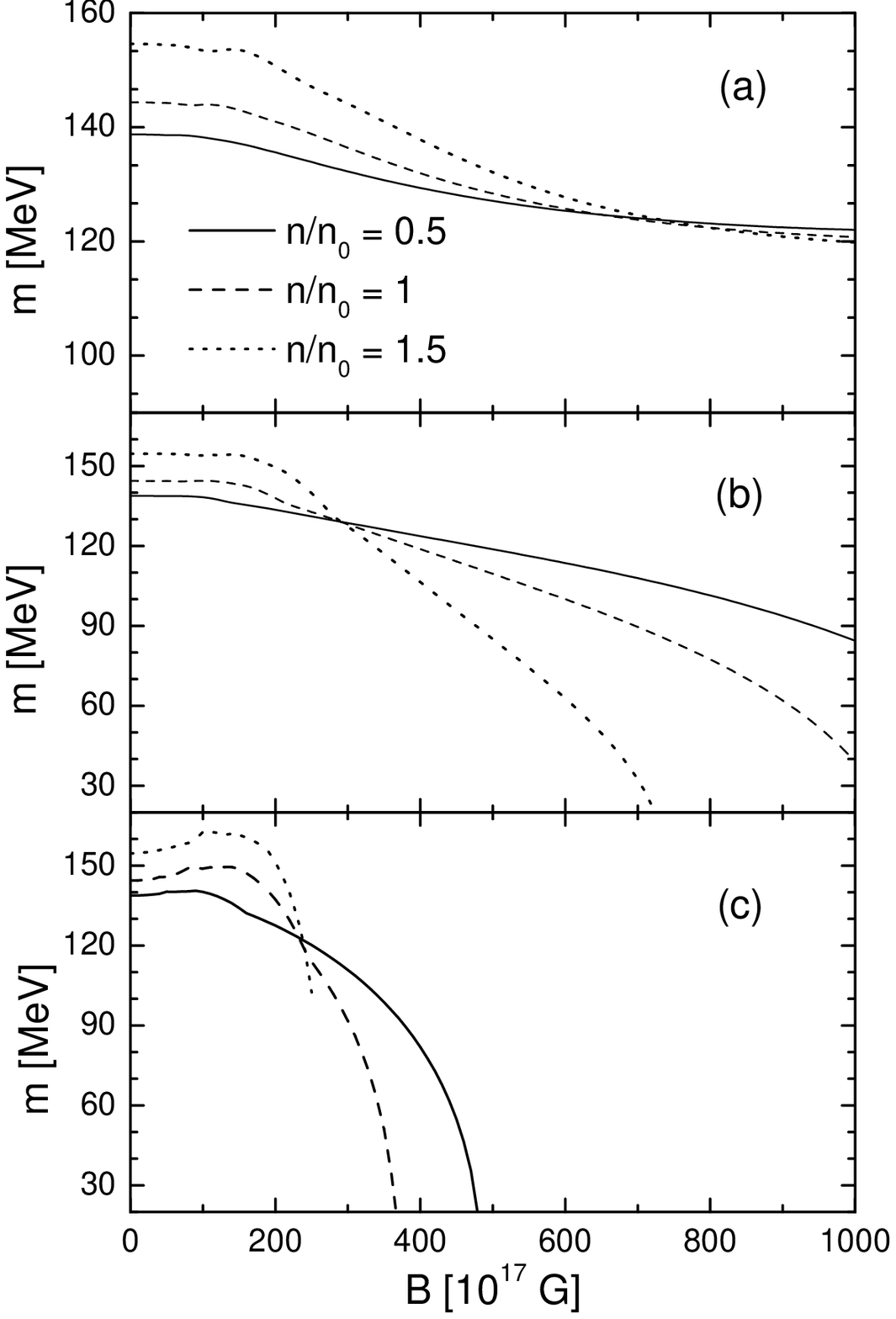}
\caption{ The effective pion mass as a function of the magnetic
intensity for three different values of the baryon density using
the sets AMM0 (a), AMM1 (b) and AMM2 (c). Isospin conservation
under the condition of flavor symmetry $n_u=n_d$ is considered. }
\end{figure}

\newpage
\begin{figure}
\includegraphics[width=0.9\textwidth]{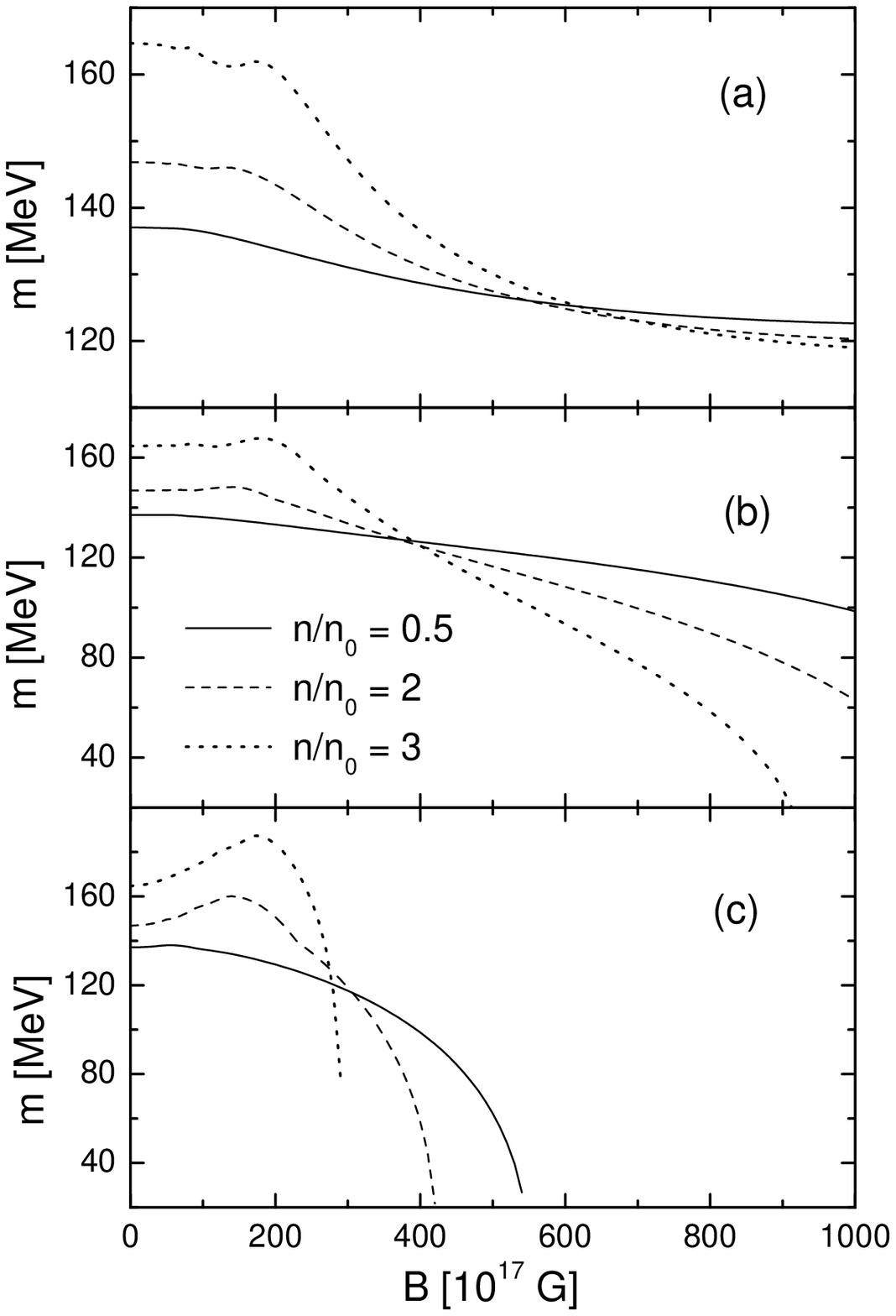}
\caption{ The effective pion mass as a function of the magnetic
intensity for three different values of the baryon density using
the sets AMM0 (a), AMM1 (b) and AMM2 (c). Isospin conservation
under the condition of electrical neutrality $2 \,n_u=n_d$ is
considered. }
\end{figure}
\newpage

 \newpage
 \begin{figure}
 \includegraphics[width=0.9\textwidth]{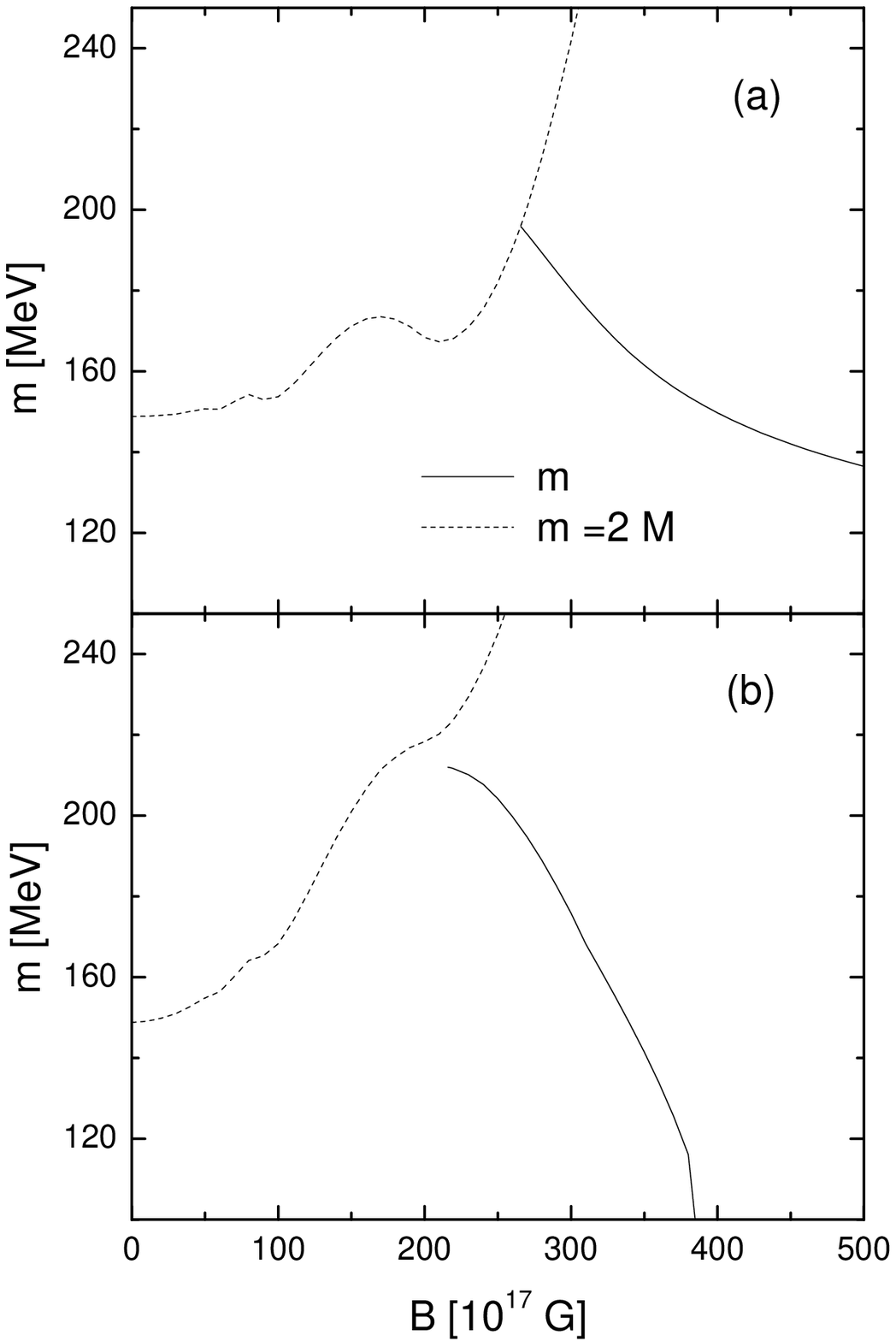}
 \caption{ The threshold condition $m=2\, M$ for the stability of the solutions shown
 in the $m-B$ plane, and the actual stable solutions of the pion mass for dense quark
 matter at $n/n_0=4$ in the $2 \,n_u=n_d$ case. Different results corresponding to the AMM0 (a)
 or the AMM1 (b) sets are shown.}
 \end{figure}

 \newpage
 \begin{figure}
 \includegraphics[width=0.9\textwidth]{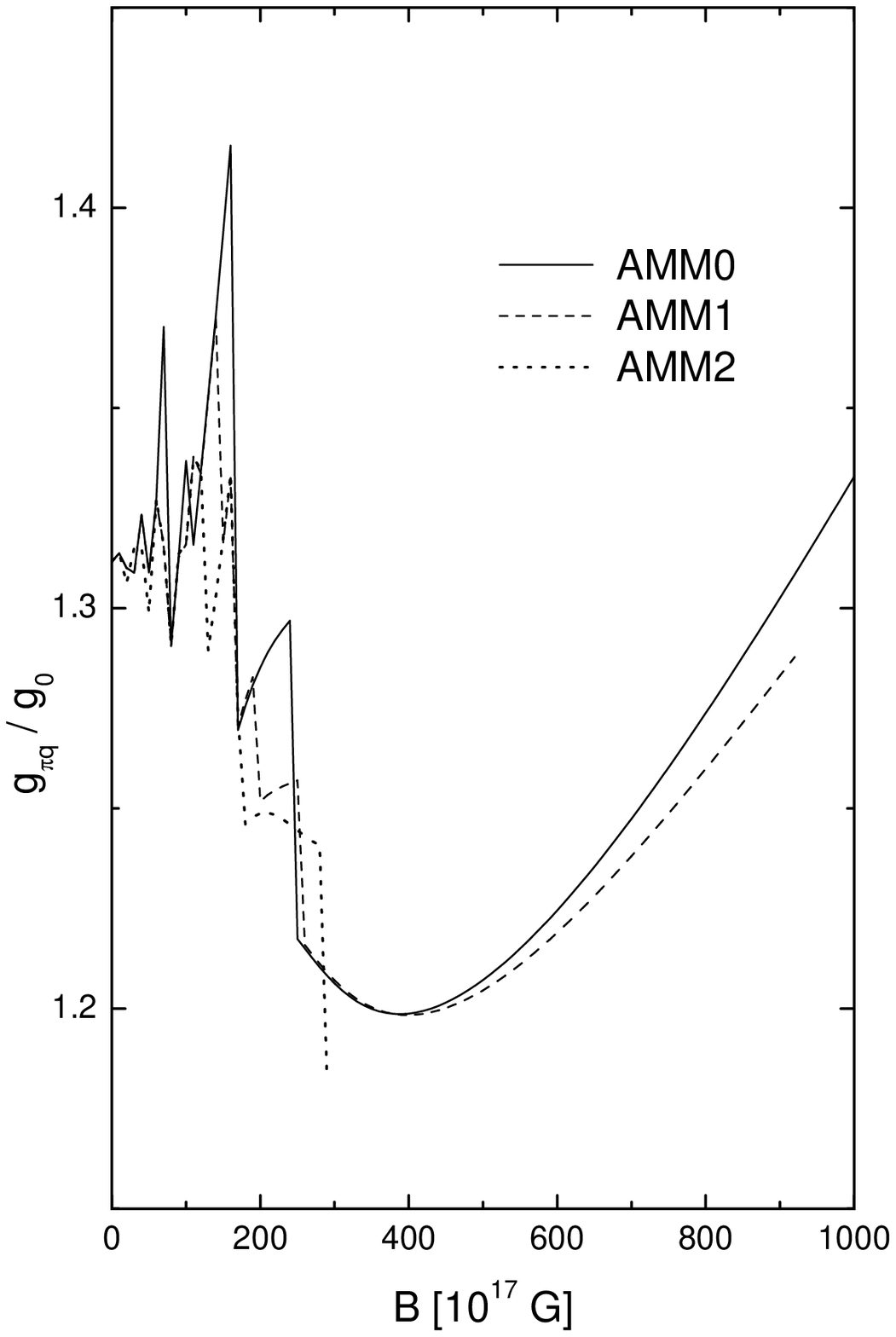}
 \caption{ The pion-quark effective coupling as a function of the magnetic intensity
 evaluated in dense quark matter at $n/n_0=3$ under the condition $2 \,n_u=n_d$. }
 \end{figure}

\end{document}